\documentclass[pra,twocolumn,showpacs,superscriptaddress,amsmath]{revtex4-1}
\usepackage{graphicx}
\usepackage{upgreek}
\usepackage{color}

\newcommand{\ket}[1]{\left|{#1}\right>}
\newcommand{\bra}[1]{\left<{#1}\right|}
\newcommand{\tr}[1]{\textnormal{tr}{\left\{#1\right\}}}
\newcommand{\I}{\mathrm{i}}

\newcommand{\ID}{D_\text{i}}
\newcommand{\OD}{D_\text{o}}

\begin{document}

\title{Adaptive schemes for incomplete quantum process tomography}

\author{Yong Siah Teo}
\affiliation{Centre for Quantum Technologies, National University of Singapore, Singapore 117543, Singapore}
\affiliation{NUS Graduate School for Integrative Sciences and Engineering, Singapore 117597, Singapore}
\author{Berthold-Georg Englert}
\affiliation{Centre for Quantum Technologies, National University of Singapore, Singapore 117543, Singapore}
\affiliation{Department of Physics, National University of Singapore, Singapore 117542, Singapore}
\author{Jaroslav {\v R}eh{\'a}{\v c}ek}
\affiliation{Department of Optics, Palacky University, 17. listopadu 12, 77146 Olomouc, Czech Republic}
\author{Zden{\v e}k Hradil}
\affiliation{Department of Optics, Palacky University, 17. listopadu 12, 77146 Olomouc, Czech Republic}

\begin{abstract}
We propose an iterative algorithm for incomplete quantum process tomography, with the help of quantum state estimation, based on the combined principles of maximum-likelihood and maximum-entropy. The algorithm yields a unique estimator for an unknown quantum process when one has less than a complete set of linearly independent measurement data to specify the quantum process uniquely. We apply this iterative algorithm adaptively in various situations and so optimize the amount of resources required to estimate the quantum process with incomplete data.
\end{abstract}

\date{\today}

\begin{widetext}
\maketitle
\end{widetext}

\section{Introduction}

Quantum process tomography (QPT) is an important tool to characterize the operation of a given quantum channel \cite{qpt0,qpt1,qpt2,qpt3}. Such a characterization is needed, for example, when one attempts to construct a quantum channel comprising multiple logic gates, each carrying out a specific quantum process. One such quantum channel for entanglement distillation, for instance, would consist of CNOT gates. A physical quantum process is described by a completely-positive map $\mathcal{M}$. That is, given a particular input quantum state $\rho_\text{i}$ residing in the $\ID$-dimensional Hilbert space $\mathcal{H}$, the resulting output state $\rho_\text{o}$ in the $\OD$-dimensional Hilbert space $\mathcal{K}$ is given by
\begin{equation}
\rho_\text{o}=\mathcal{M}\left(\rho_\text{i}\right)=\sum_{m}K_m\rho_\text{i}K^\dagger_m\,,
\label{cptp}
\end{equation}
with the Kraus operators $K_m$ satisfying the relation $\sum_mK^\dagger_mK_m=1_\mathcal{K}$. The $K_m$s are not unique and any other set of Kraus operators
\begin{equation}
K'_m=\sum_{m'}u_{m'm}K_{m'}\,,
\end{equation}
where the $u_{m'm}$s are the elements of a unitary matrix, also parameterizes the completely-positive map $\mathcal{M}$ \cite{nielsenchuang}.

The idea behind QPT is to estimate such completely-positive maps with measurements. Much like quantum state tomography, the estimation of an unknown quantum process can be perceived as the estimation of a positive \emph{Choi-Jami{\'o}\l kowski operator} $E_\text{true}$ that is represented by a $\ID\OD\times\ID\OD$ matrix \cite{choijam}. Such an operator contains all accessible information about the quantum process. The standard QPT procedure involves the measurement of multiple copies of $L$ different output states, with each output state corresponding to one of the $L$ linearly independent input states $\rho^{(l)}_\text{i}$, thereby using a probability operator measurement (POM) of, say, $M$ outcomes. The unknown operator $E_\text{true}$ is estimated by linear-inversion of the $LM$ measurement frequencies, which consists of $\ID^2\OD^2$ linearly independent constraints. Like the linear-inversion procedure for quantum state estimation, the resulting estimator obtained may not be positive. If that is the case, the estimator cannot be used for statistical predictions. This failure occurs whenever the observed relative frequencies of the measurement outcomes do not have consistent interpretation as probabilities. What is, therefore, called for, is an estimation procedure that ensures a physically meaningful estimator whatever the measurement data may be.

One statistically meaningful technique to obtain a positive estimator for $E_\text{true}$ is the maximum-likelihood estimation procedure (ML) \cite{rehacek1}. This can be applied to yield a unique estimator $\hat E_\text{ML}$ as long as the measurement data obtained form a set of $\ID^2\OD^2$ linearly independent constraints. We say that this set of measurement data is informationally complete. However, the number of linearly independent parameters increases rapidly with the dimensions and a complete characterization of $E_\text{true}$ becomes unfeasible for complex processes. This is especially true when the quantum process acts on an infinite-dimensional Hilbert space \cite{csqpt}. The well-known method of \emph{Direct Characterization of Quantum Dynamics} (DCQD) \cite{qpt1} was introduced to reduce the amount of measurement resources that are used for quantum process tomography. However, this method requires entangled input states and post-processing strategies that can be expensive when dealing with more complex quantum processes.

A more straightforward and conceptually different approach is to resort to informationally incomplete QPT. With this approach, less measurement resources are used to obtain an estimator for the unknown quantum process to a fair amount of accuracy. As a consequence, there exists a convex set of infinitely many ML estimators which are consistent with the measurement data. To choose the estimator which is least-biased from the convex set, we invoke the maximum-entropy principle \cite{jaynes} and choose the estimator with the largest entropy. Such an incomplete QPT can also give useful information about the quantum channel. In a typical tomography experiment, with data from measuring a finite number of copies, the resulting quantum process estimator can never be exactly equal to $E_\text{true}$ since experimental fluctuations are inevitable. One can only obtain an estimator that is close to $E_\text{true}$ within a certain tomographic precision. Thus, MLME QPT is typically useful in providing a unique estimator for an unknown quantum process within a suitable tomographic precision using fewer incomplete measurement resources. As will be shown, this reduction in measurement resources is more pronounced for unitary quantum channels. Since $E_\text{true}$ is unknown, one common practice is to gauge such a tomographic precision with another operator $E_\text{prior}$ that is close to $E_\text{true}$, based on some prior information one has about the constructed quantum channel. The availability of such a $E_\text{prior}$ for a given $E_\text{true}$ will become useful and important in subsequent discussions.

The estimators obtained using the aforementioned method are least-biased with respect to the set of incomplete measurement data in the sense of the \emph{entropy of the quantum process}. In Ref.~\cite{iqpt}, which is an analytical study of the conventional maximum-entropy method, the entropy functional for the Choi-Jami{\'o}{\l}kowski operator $E$ describing a quantum channel was introduced as $S\left(E\right)=-\tr{(E/\ID)\log (E/\ID)}$ and this was shown to exhibit nice properties. In particular, this concave channel entropy functional has a unique maximum in $E$ and is zero only when the quantum channel is unitary since $E/\ID$ is then a rank-1 projector. However, the analytical results in \cite{iqpt} apply only to simple qubit channels and are difficult to extend to general quantum channels of greater complexity.

In this article, we shall extend the strategy in Ref.~\cite{mlme} and establish an adaptive iterative algorithm to search for the MLME estimator $\hat E_\text{MLME}$ which maximizes both the likelihood and entropy functionals using the channel entropy functional in \cite{iqpt}. We first give some preliminary ideas on quantum process estimation in Sec.~\ref{sec:prelim}. Then, in Sec.~\ref{sec:iteralgo}, we will present the iterative MLME algorithm using variational principles to derive a steepest-ascent scheme and apply it to numerical simulations of two-qubit and three-qubit quantum channels. In Sec.~\ref{sec:adaptstrat}, we will establish adaptive strategies to apply the MLME algorithm with the aim of minimizing the amount of measurement resources needed to perform incomplete QPT.

\section{Preliminaries of quantum process estimation}
\label{sec:prelim}

The estimation of the completely-positive map $\mathcal{M}$ that describes an unknown quantum process, in the manner presented in Eq.~(\ref{cptp}), is isomorphic to the estimation of an unknown quantum state. This is a consequence of the well-known Choi-Jami{\'o}{\l}kowski isomorphism \cite{choijam,rehacek1}. Let us define a maximally-entangled pure state $\ket{\Psi_+}=\sum_j\ket{j}_\mathcal{H}\otimes\ket{j}_\mathcal{H'}/\sqrt{\ID}$ in terms of the computational basis kets $\ket{j}_\mathcal{H}\otimes\ket{j}_\mathcal{H'}$. Here, the dimensions of the Hilbert spaces $\mathcal{H}$ and $\mathcal{H'}$ are both equal to the dimension $\ID$ of the input Hilbert space. Using this basis, there exists a one-to-one correspondence between the map $\mathcal{M}$ and a unique positive operator $E$ defined as follows:
\begin{align}
E\,\equiv&\ID\left(\mathcal{I}_\mathcal{H}\otimes\mathcal{E}_\mathcal{H'}\right)\left(\ket{\Psi_+}\bra{\Psi_+}\right)\,\\
\widehat{=}&\sum_{jk}\left(\ket{j}\bra{k}\right)\otimes\mathcal{M}\left(\ket{j}\bra{k}\right)\,,
\end{align}
with $\mathcal{I}_\mathcal{H}$ being the identity map. From Eq.~(\ref{cptp}), the alternative expression
\begin{equation}
E=\sum_m\ket{\psi_m}\bra{\psi_m}\,,
\end{equation}
with
\begin{equation}
\quad\ket{\psi_m}=(1_\mathcal{H}\otimes K_m)\ket{\Psi_+}\sqrt{\ID}\,,
\end{equation}
implies that the rank of $E$ is equal to the number of linearly independent $K_m$s. It follows that $E$ is rank-1 if the completely-positive map is described by a single unitary Kraus operator, and only then.

The output state can be expressed in terms of $E$ by means of
\begin{equation}
\rho_\text{o}=\mathrm{tr}_\mathcal{H}\left\{E\left(\rho_\text{i}^\mathrm{T}\otimes 1_\mathcal{K}\right)\right\}\,,
\end{equation}
where the transposition is defined with respect to the computational basis. Hence, reconstructing the quantum process amounts to estimating the positive operator $E$. To do so, one requires a total of $\ID^2\OD^2$ real parameters to specify the corresponding matrix. In the subsequent analyses, we shall consider trace-preserving maps, that is $\tr{\rho_\text{i}}=\tr{\rho_\text{o}}$ for any $\rho_\text{i}$, in which case the number of independent parameters is reduced to $\ID^2(\OD^2-1)$, with the constraints compactly written as
\begin{equation}
\mathrm{tr}_\mathcal{K}\left\{E\right\}=1_\mathcal{H}\,.
\label{constraints}
\end{equation}

To estimate $E$, typically a set of $L$ input states $\rho^{(l)}_\text{i}$, with $N$ copies each, are sent through the quantum channel, one state at a time. The output state $\rho^{(l)}_\text{o}$ that corresponds to $\rho^{(l)}_\text{i}$ is measured with a POM consisting of $M$ outcomes $\Pi_m\geq0$ such that $\sum_m\Pi_m=1_\mathcal{K}$. The probability of getting outcome $\Pi_m$ for the input state $\rho^{(l)}_\text{i}$ is given by $p_{lm}=\tr{E\left(\rho^{(l)\,\mathrm{T}}_\text{i}\otimes\Pi_m\right)}{\Big/}L$. Here, $p'_l\equiv\sum_mp_{lm}=1/L$. If the $LM$ parameters comprise $\ID^2\OD^2$ linearly independent ones, the measurement data will be informationally complete. One can thus perform a complete quantum process estimation using the maximum-likelihood (ML) algorithm \cite{rehacek1} and so obtain a unique positive estimator $\hat E_\text{ML}$ by maximizing the likelihood functional
\begin{equation}
\mathcal{L}(E)=\prod^L_{l=1}\left(\prod^M_{m=1}p_{lm}^{n_{lm}}\right)\,,
\end{equation}
where the number of occurrences $n_{lm}$ for the outcome $\Pi_m$ obtained in an experiment with the input state $\rho^{(l)}_\text{i}$ are such that $n'_l\equiv\sum_{m}n_{lm}=N$.

In what follows, we shall discuss the situation in which one performs quantum process estimation with a set of informationally incomplete measurement data. All processes considered henceforth will be trace-preserving, but a generalization to those which are not trace-preserving is straightforward.

\section{The iterative algorithm}
\label{sec:iteralgo}

We consider the optimization of the information functional
\begin{equation}
I(\lambda;E)=\lambda S(E)+\frac{1}{LN}\log\mathcal{L}(E)\,,
\label{newinfo}
\end{equation}
where $\lambda$ is a parameter which scales the entropy relative to the normalized log-likelihood and should be chosen with a very small value. When the measurement data are informationally complete, one sets $\lambda$ to zero and optimizing $I(\lambda=0;E)$ amounts to the ML problem \cite{rehacek1}. In the same spirit as in \cite{mlme}, both our knowledge from the measurement data (contained in $\log\left(\mathcal{L}(E)\right)/LN$ which measures the information gain) and our ignorance (reflected in $S(E)$ which measures the lack of information) about the operator $E$ are taken into account in such a way that our ignorance takes an infinitesimal weight. This introduces a small and smooth convex hill over the set of positive ML estimators which selects the one with the largest entropy. As in \cite{mlme}, the value of $\lambda$ may be chosen such that both $\log\left(\mathcal{L}(E)\right)/LN$ and $S(E)$ remain almost constant with respect to $\lambda$.

To maximize $I(\lambda;E)$ with respect to $E$, we define the variation $E+\updelta E=(1+Z^\dagger)E(1+Z)$, where $Z$ is a small arbitrary operator such that Eq.~(\ref{constraints}) is satisfied, that is: $\tr{\updelta E}=0$. Thus the most general expression for $Z$ is
\begin{equation}
1+Z=\left(1+\updelta A\right)\left[\sqrt{\mathrm{tr}_\mathcal{K}\left\{\left(1+\updelta A^\dagger\right)E\left(1+\updelta A\right)\right\}}\otimes 1_\mathcal{K}\right]^{-1}\,,
\label{variations}
\end{equation}
with an unrestricted infinitesimal $\updelta A$. On the other hand, the variation of $I(\lambda;E)$ with respect to $E$ gives $\tr{\updelta E\,W}$, where
\begin{equation}
W=\frac{1}{L}\sum_{lm}\frac{f_{lm}}{p_{lm}}\rho^{(l)\,\mathrm{T}}_\text{i}\otimes\Pi_m-\frac{\lambda}{\ID}\left[1+\log\left(\frac{E}{\ID}\right)\right]\,
\label{auxop}
\end{equation}
and $f_{lm}=n_{lm}/LN$.
Keeping only first-order variations systematically and imposing $\updelta I(\lambda;E)>0$, the method of steepest ascent leads us to
\begin{equation}
\updelta A=\updelta A^\dagger=\frac{\epsilon}{2}\left(W-\frac{1}{2}\mathrm{tr}_\mathcal{K}\left\{WE+EW\right\}\otimes 1_\mathcal{K}\right)
\end{equation}
for some small $\epsilon>0$. Hence, to obtain the MLME estimator $\hat E_\text{MLME}$, one simply fixes $\lambda\ll 1$ and iterates the equations
\begin{align}
E_{n+1}=&\,(1+Z^\dagger_n)E_n(1+Z_n)\,,\nonumber\\
\left(\updelta A\right)_n=&\,\frac{\epsilon}{2}\left(W_n-\frac{1}{2}\mathrm{tr}_\mathcal{K}\left\{W_n E_n+E_n W_n\right\}\otimes 1_\mathcal{K}\right)\,,\label{iter1}
\end{align}
where the expression for $Z_n$ follows from Eq.~(\ref{variations}) and $W_n$ denotes the operator $W$ in Eq.~(\ref{auxop}) evaluated for $E_n$. One may do so by starting from a randomly chosen operator $E_0$ and continue until the extremal equation for $\hat E_\text{MLME}$ is satisfied with some pre-chosen numerical precision. To derive this extremal equation, we define the Lagrange functional \cite{rehacek1}
\begin{equation}
\mathcal{D}(E)=I(\lambda;E)-\tr{\Lambda\,E}
\end{equation}
with the Lagrange operator $\Lambda\equiv h\otimes 1_\mathcal{K}$ for the constraints in Eq.~(\ref{constraints}), where $h$ is a Hermitian operator. Setting the variation of $\mathcal{D}(E)$ to zero gives the extremal equation
\begin{equation}
\Lambda\hat E_\text{MLME}\Lambda=W_\text{MLME}\hat E_\text{MLME}W_\text{MLME}
\end{equation}
with $\Lambda=\sqrt{\mathrm{tr}_\mathcal{K}\left\{W_\text{MLME}\hat E_\text{MLME}W_\text{MLME}\right\}}\otimes 1_\mathcal{K}$.

Thus far, we have been assuming that the measurement outcomes $\Pi_m$ give perfect detection of quantum systems. The iterative equations in Eq.~(\ref{iter1}) can be generalized to the case of imperfect detection. In this case it is clear that, if each of the $M$ measurement outcomes $\Pi_m$ is assigned a detection efficiency $\eta_m\leq1$, one can always define a new set of $M$ measurement outcomes $\tilde\Pi_m\equiv\eta_m\Pi_m$ such that $G\equiv\sum_m\tilde\Pi_m\neq1_\mathcal{K}$. The iteration procedure of Eq.~(\ref{iter1}) can still be used with the new set of POM outcomes $\tilde\Pi_m$ provided that the operator $W$ in Eq.~(\ref{auxop}) is replaced by $W-W_0$, where
\begin{equation}
W_0=\frac{1}{L\sum_{l'}p'_{l'}}\sum_l\rho^{(l)\,\mathrm{T}}_\text{i}\otimes G\,
\end{equation}
accounts for the copies that escape detection.

As an example, we apply the algorithm to numerical simulations on two-qubit channels, the CNOT gate described by the unitary operator
\begin{equation}
U_\text{CNOT}\,\widehat{=}\begin{pmatrix}
1 & 0 & 0 & 0\\
0 & 1 & 0 & 0\\
0 & 0 & 0 & 1\\
0 & 0 & 1 & 0
\end{pmatrix}
\end{equation}
and a randomly generated non-unitary quantum channel described by a full-rank Choi-Jami{\'o}{\l}kowski matrix, as well as the three-qubit Toffoli gate described by the unitary operator
\begin{equation}
U_\text{Toffoli}\,\widehat{=}\begin{pmatrix}
1 & 0 & 0 & 0 & 0 & 0 & 0 &0\\
0 & 1 & 0 & 0 & 0 & 0 & 0 &0\\
0 & 0 & 1 & 0 & 0 & 0 & 0 &0\\
0 & 0 & 0 & 1 & 0 & 0 & 0 &0\\
0 & 0 & 0 & 0 & 1 & 0 & 0 &0\\
0 & 0 & 0 & 0 & 0 & 1 & 0 &0\\
0 & 0 & 0 & 0 & 0 & 0 & 0 &1\\
0 & 0 & 0 & 0 & 0 & 0 & 1 &0\\
\end{pmatrix}\,.
\end{equation}
To quantify the discrepancy between an MLME estimator and the true Choi-Jami{\'o}{\l}kowski operator $E_\text{true}$, we use the trace-class distance
\begin{equation}
\mathcal{D}_\text{tr}\left(\hat E_\text{MLME},E_\text{true}\right)=\frac{1}{2\ID}\tr{\big|\hat E_\text{MLME}-E_\text{true}\big|}\,,
\end{equation}
where $|A|=\sqrt{A^\dagger A}$ for any operator $A$. In these simulations, we take the $\ID$-dimensional projectors of a symmetrically informationally complete POM (SIC POM) as the input states \cite{sicpom}. One such set of states contains the minimal number of $\ID^2$ pure states $\ket{\psi_j}\bra{\psi_j}$ such that
\begin{equation}
\left\vert\langle\psi_j|\psi_k\rangle\right\vert^2=\frac{\ID\delta_{jk}+1}{\ID+1}\,.
\label{sicpom}
\end{equation}
\begin{figure}[htp]
  \centering
  \includegraphics[width=0.4\textwidth]{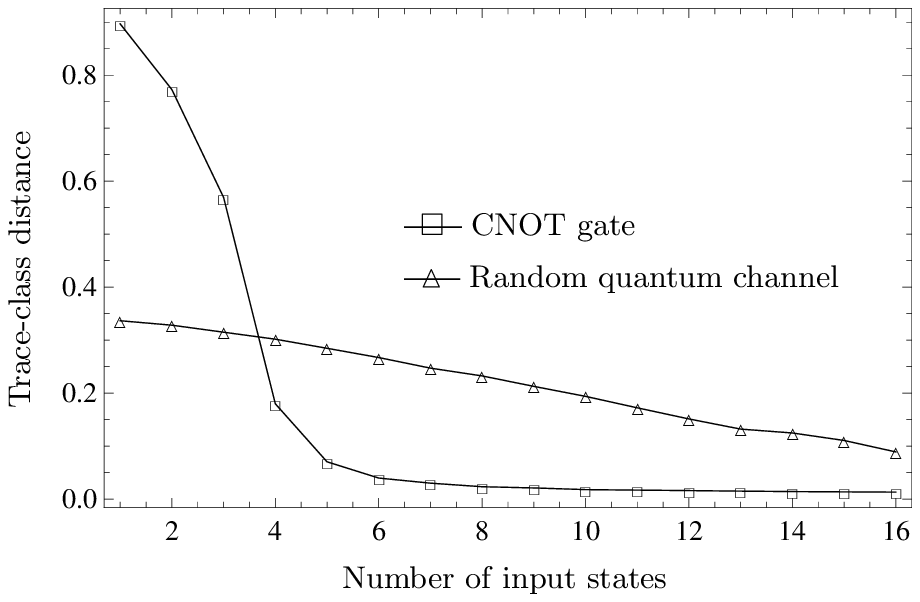}\\
  \includegraphics[width=0.4\textwidth]{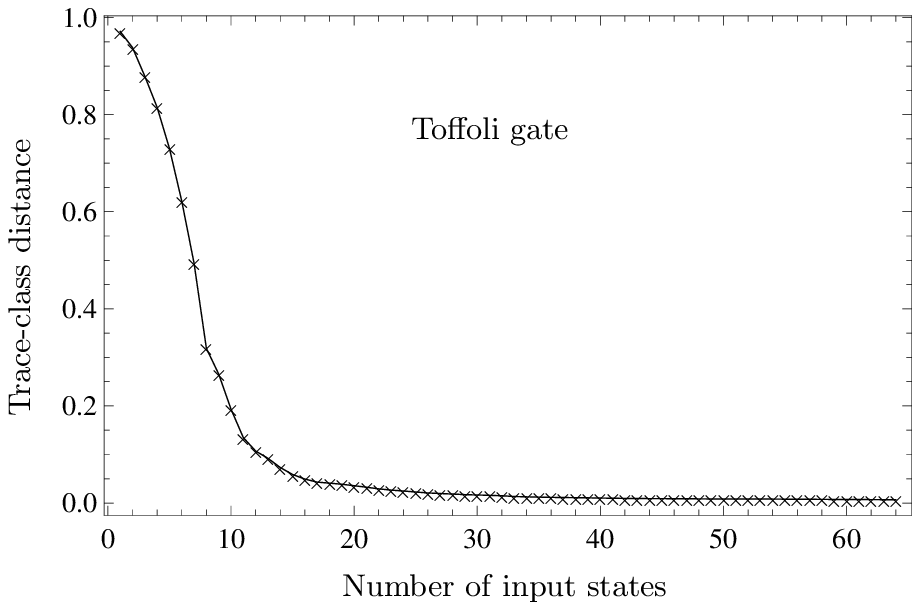}
  \caption{Numerical simulations on the two-qubit ($d=2^2$) and three-qubit ($d=2^3$) quantum channels where $\ID=\OD=d$. The projectors of symmetric informationally complete POMs (SIC POMs) are chosen as the linearly independent input states for all the simulations ($L=d^2$). For the measurements, informationally complete POMs consisting of tensor products of qubit SIC POMs are used ($M=d^2$). Each qubit SIC POM consists of a set of pure states whose Bloch vectors form the ``legs of a tetrahedron'' in the Bloch sphere. For the two-qubit channels, $N=10^4$ and an average over 50 experiments is taken to compute the trace-class distances. For the three-qubit channel, the measurement data are generated without statistical noise. For unitary channels, one can see that the MLME algorithm can still give fairly accurate estimations with a smaller number of input states than that of a linearly independent set. Numerical simulations of arbitrary two-qubit and three-qubit unitary channels suggest that the number is approximately $d^2/2$ for SIC POM input states, above which there is insignificant tomographic improvement.}
  \label{fig:channels}
\end{figure}As shown in Fig.~\ref{fig:channels}, using the MLME algorithm for QPT can give fast convergence in terms of tomographic efficiency with a reduced number of input states as quantum resources. This reduction is especially significant for unitary processes, where the Choi-Jami{\'o}{\l}kowski operators are rank-1. For nonunitary quantum processes described by matrices of larger rank, the tomographic efficiency will be lower as shown in the first plot of Fig.~\ref{fig:channels}. This is expected in analogy with quantum state tomography where it is more difficult to reliably estimate highly-mixed states than nearly-pure ones.

\section{Adaptive strategies}
\label{sec:adaptstrat}

An interesting question to ask with regard to incomplete QPT is whether one can perform it in an optimal way given the available resources by means of adaptive strategies. Here optimality refers to the minimization of the amount of resources (input states or measurements) used to perform incomplete QPT such that the distance between $\hat E_\text{MLME}$ and $E_\text{true}$ reaches a certain desired value. Very frequently, despite the fact that $E_\text{true}$ is always unknown, one has a rough idea of an operator $E_\text{prior}$ which may be close to $E_\text{true}$ based on some prior information about the unknown $E_\text{true}$. This scenario is reasonable and typical when one designs a quantum channel experimentally which performs an expected quantum operation, with errors arising from imperfections of the components that make up the channel. We shall establish adaptive strategies which make use of such an operator in order to select, with the help of the MLME algorithm, resources for incomplete QPT in an optimal way. We refer to such tomography schemes as the \emph{adaptive MLME quantum process tomography} (AMLME QPT) schemes.

We will focus on adaptive strategies to choose the input states optimally. This can be reviewed in two separate cases: The case in which a fixed set of linearly independent input states is used (Sec.~\ref{subsec:fixed_input}) and that in which arbitrary input states can be generated for incomplete QPT (Sec.~\ref{subsec:all_input}). The optimization of the POM will be studied on a later occasion.

\subsection{Optimization over a fixed set of linearly independent input states}
\label{subsec:fixed_input}
In the previous section, we considered the projectors of the SIC POMs, which are known to have optimal tomographic efficiencies, as input states in the numerical simulations. Since these POMs are symmetric in the sense of Eq.~(\ref{sicpom}), any ordering of the input states in a given set gives the same plots in Fig.~\ref{fig:channels}. In practice, however, such entangled states are difficult to produce and one typically has access to a set of separable states \cite{iontrap} for measurements instead. In this case, there no longer exists such a symmetry and the tomographic performance depends on the order of the input states chosen, possibly strongly so. We propose to optimize the tomographic performance by choosing the input states adaptively based on the measurement data collected from the previously chosen input states, thereby using the prior $E_\text{prior}$.

To describe the adaptive strategy, let us consider a set of $L\geq D^2_\text{i}$ input states in which $D^2_\text{i}$ of them are linearly independent. Suppose that $N$, which is a fixed integer for all input states, copies of a randomly chosen input state $\rho_\text{i}^{(1)}$ are sent through the quantum channel and the first set of measurement data $\{\nu_{11},\ldots,\nu_{1M}\}$, $\sum_m\nu_{1m}=1$, is collected. With these data $f_{1m}\equiv\nu_{1m}$, one obtains the first MLME estimator $\hat E^{(1)}_\text{MLME}$. To select the next input state out of the remaining $L-1$ states, we take $E_\text{prior}$ as a gauge for $E_\text{true}$ to generate $L-1$ sets of probabilities respectively from the $L-1$ states. Each set of probabilities is then treated as the set of frequencies $\{\nu^{(k)}_{21},\ldots,\nu^{(k)}_{2M}\}$, for the corresponding input state $k$. Hence, one has $L-1$ sets of measurement data, each set being the combined data $\{\nu_{11},\ldots,\nu_{1M},\nu^{(k)}_{21},\ldots,\nu^{(k)}_{2M}\}/2$ with the normalized frequencies $f_{1m}\equiv\nu_{1m}/2$ and $f^{(k)}_{2m}\equiv\nu^{(k)}_{2m}/2$ such that $\sum_m(f_{1m}+f^{(k)}_{2m})=1$ for each $k$, and the corresponding $L-1$ \emph{projected} MLME estimators $\hat E^{(2)}_{\text{MLME},k}$.

The value of $k$ is selected such that a chosen figure of merit which quantifies the distance between $\hat E^{(2)}_{\text{MLME},k}$ and $\hat E^{(1)}_\text{MLME}$ is the largest, so that there is a high chance for the next MLME estimator to be closer to $E_\text{true}$. As an example, the figure of merit is taken to be the trace-class distance $\mathcal{D}_\text{tr}\left(\hat E^{(2)}_{\text{MLME},k},\hat E^{(1)}_\text{MLME}\right)$. With this input state, the second estimator $\hat E^{(2)}_\text{MLME}$ is then obtained with MLME QPT. One repeats this procedure for subsequent input states until the distance $\mathcal{D}_\text{tr}\left(\hat E^{(l+1)}_\text{MLME},\hat E^{(l)}_\text{MLME}\right)$ is below some preset threshold. An alternative to this would be to \emph{minimize} the trace-class distance $\mathcal{D}_\text{tr}\left(\hat E^{(l+1)}_{\text{MLME},k},E_\text{prior}\right)$.

It is important to understand that in this strategy, the prior information $E_\text{prior}$ is \emph{not} used to reconstruct the unknown quantum process in any way. It serves only as a means to optimally select the input states from the given set so as to maximize the tomographic convergence. This adaptive strategy also relies partially on the measurement data obtained in the experiment. We have thus introduced an operational method of using the prior information to minimize the amount of resources needed to perform reliable MLME QPT without introducing any artifacts coming from the prior information into the reconstruction procedure. To summarize, the adaptive MLME strategy is as follows:
\begin{enumerate}
  \item Randomly choose $\rho_\text{i}^{(1)}$ from the set of $L$ input states and set $l=1$.
  \begin{enumerate}
    \item Perform QPT using $\rho_\text{i}^{(l)}$ and obtain the set of frequencies $\{\nu_{l1},\ldots,\nu_{lM}\}$, $\sum_m\nu_{lm}=1$.
    \item Set $\nu=\bigcup^l_{j=1}\{\nu_{j1},\ldots,\nu_{jM}\}/\,l$.
    \item Invoke the MLME algorithm with $\nu$ and obtain $\hat E^{(l)}_\text{MLME}$. Use $E_\text{prior}$ to compute the frequencies $\{\nu^{(k)}_{l+1\,1},\ldots,\nu^{(k)}_{l+1\,M}\}$, $\sum_m\nu^{(k)}_{l+1\,m}=1$, from the remaining input states, with $k$ labeling the remaining $L-l$ states.
    \item Define $L-l$ sets of accumulated frequencies \mbox{$(\nu\cup\{\nu^{(k)}_{l+1\,1},\ldots,\nu^{(k)}_{l+1\,M}\})/(l+1)$} and calculate the $L-l$ projected MLME estimators $\hat E^{(l+1)}_{\text{MLME},k}$.
    \item Set $\rho_\text{i}^{(l+1)}$ as the input state corresponding to $k$ such that $\mathcal{D}_\text{tr}\left(\hat E^{(l+1)}_{\text{MLME},k},\hat E^{(l)}_\text{MLME}\right)$ is largest.
  \end{enumerate}
  \item Set $l=l+1$ and repeat Steps 1(a)--1(e).\\
\end{enumerate}

\subsection{Optimization over the Hilbert space}
\label{subsec:all_input}

More generally, the adaptive strategy may be extended to the case in which one has access to the entire Hilbert space of states. In other words, the task is to search for the next optimal input state $\rho^{(L+1)}_\text{i}$ from the $\ID$-dimensional Hilbert space based on the measurement data $\nu_{lm}$ obtained in the experiment from $L$ previously chosen input states, where $\sum_m\nu_{lm}=1$ for all $l$, and the prior information $E_\text{prior}$ about the unknown quantum process.

To this end, we define the normalized \emph{projected} log-likelihood functional
\begin{equation}
\log\tilde{\mathcal{L}}(E,\rho)=\sum_{lm}\frac{\nu_{lm}}{L+1}\log\left(\tilde p_{lm}\right)+\sum_m\frac{\tilde \nu_m}{L+1}\log\left(\tilde p_m\right)\,,
\label{crossent}
\end{equation}
where
\begin{center}
\begin{equation*}
\tilde p_{lm}=\tr{E\,\frac{\rho^{(l)\,T}_\text{i}\otimes\Pi_m}{L+1}}\,,
\end{equation*}
\end{center}
\begin{equation*}
\tilde \nu_m=\tr{E_\text{prior}\rho^T\otimes\Pi_m}\,\,\,\text{and}\,\,\,\tilde p_m=\tr{E\,\frac{\rho^T\otimes\Pi_m}{L+1}}\,
\end{equation*}
with $l$ always running from $1$ to $L$ over all previously used input state labels. This projected log-likelihood functional is a good approximation to the log-likelihood functional for the situation in which the state $\rho$ is chosen as the next input state for the experiment as long as $E_\text{prior}$ is not too far away from $E_\text{true}$. The projected frequencies $\tilde \nu_m$ estimate the actual frequencies one gets when measuring the input state $\rho$. An optimal input state $\rho^{(L+1)}_\text{i}$ and the corresponding Choi-Jami{\'o}{\l}kowski operator are chosen as the positive estimators that maximize this projected log-likelihood functional.

Coincidentally, this maximum projected log-likelihood (MPL) procedure is equivalent to minimizing the cross entropy functional $\mathcal{C}(E,\rho)=-\log\tilde{\mathcal{L}}(E,\rho)$ \cite{mce1,mce2} over all positive operators subjected to the respective constraints for $\rho$ and $E$. Hence, another way of understanding this procedure is to first regard both the incomplete measurement data collected after using $L$ input states and $E_\text{prior}$ as the full prior knowledge one has about the unknown $E_\text{true}$. The statistical motivation for MPL or minimizing $\mathcal{C}(E,\rho)$ is, loosely speaking, to obtain estimators which are as compatible with this prior knowledge as possible by minimizing the entropy of the prior knowledge $\mathcal{C}(E,\rho)$. We will provide some more arguments related to this optimization technique in the later part of this section.

To carry out the optimization, we consider the response of $\log\tilde{\mathcal{L}}(E,\rho)$ to small variations of both $\rho$ and $E$. After some similar calculations as in Sec.~\ref{sec:iteralgo}, we obtain the MPL iterative equations
\begin{align}
E_{n+1}=&\,(1+Z^\dagger_n)E_n(1+Z_n)\,,\nonumber\\
\rho_{n+1}=&\,\frac{(1+\epsilon_2\Xi_n)\rho_n(1+\epsilon_2\Xi_n)}{\text{tr}_\mathcal{H}\{(1+\epsilon_2\Xi_n)\rho_n(1+\epsilon_2\Xi_n)\}}\,,
\label{iter2}
\end{align}
where $Z_n$ is defined by Eq.~(\ref{variations}) with
\begin{align}
\left(\updelta A\right)_n=&\,\frac{\epsilon_1}{2}\left(X_n-\frac{1}{2}\mathrm{tr}_\mathcal{K}\left\{X_n E_n+E_n X_n\right\}\otimes 1_\mathcal{K}\right)\,,\nonumber\\
X_n=&\,\sum_{lm}\frac{\nu_{lm}}{\tilde p_{lm}}\frac{\rho^{(l)\,T}_\text{i}\otimes\Pi_m}{(L+1)^2}+\sum_m\frac{\tilde \nu_m}{\tilde p_m}\frac{\rho^T\otimes\Pi_m}{(L+1)^2}\,,
\label{xop}
\end{align}
and
\begin{align}
\Xi_n\equiv &\, Y_n-\text{tr}_\mathcal{H}\{Y_n\rho_n\}\,,\nonumber\\
Y_n=&\,\text{tr}_\mathcal{K}\Bigg\{\Bigg[\sum_m\frac{1_\mathcal{H}\otimes\Pi_m}{L+1}\nonumber\\
&\,\times\left(\log\left(\tilde p_m\right)E_\text{prior}+\frac{\tilde \nu_m}{(L+1)\tilde p_m}E\right)\Bigg]^T\Bigg\}\,.
\label{yop}
\end{align}
The MPL estimators satisfy the extremal equations
\begin{align}
\tilde\Lambda\hat E_\text{MPL}\tilde\Lambda&=X_\text{MPL}\hat E_\text{MPL}X_\text{MPL}\,,\nonumber\\
\hat\rho_\text{MPL}Y_\text{MPL}&=Y_\text{MPL}\hat\rho_\text{MPL}=\text{tr}_\mathcal{H}\{Y_\text{MPL}\hat\rho_\text{MPL}\}\hat\rho_\text{MPL}\,,
\label{MPLextremal}
\end{align}
where
\begin{equation}
\tilde\Lambda=\sqrt{\mathrm{tr}_\mathcal{K}\left\{X_\text{MPL}\hat E_\text{MPL}X_\text{MPL}\right\}}\otimes 1_\mathcal{K}\,.
\label{MPLextremal2}
\end{equation}
The small parameters $\epsilon_1$ and $\epsilon_2$ are positive numbers. Thus, to carry out the MPL procedure, one iterates Eqs.~(\ref{iter2}) until Eqs.~(\ref{MPLextremal}) are satisfied with a preset numerical precision.

There is one important feature of this optimization scheme. From Eq.~(\ref{crossent}), we note that $\log\tilde{\mathcal{L}}(E,\rho)$ is neither convex nor concave in $\rho$ and hence can have multiple local maxima. Thus, the MPL optimization is non-convex. To generate these local-maxima estimators, one can start from multiple randomly chosen starting points and perform the iterations. Thereafter, the state estimator $\hat\rho_\text{MPL}$ to be chosen as the next input state $\rho^{(L+1)}_\text{i}$ is such that its corresponding $\hat E_\text{MPL}$ gives the largest trace-class distance away from the previous MLME estimator $\hat E^{(L)}_\text{MLME}$, which is obtained from the data of the previously chosen $L$ input states, over all generated pairs of MPL estimators $(\hat\rho_\text{MPL},\,\hat E_\text{MPL})$. Again, one may also minimize the distance between $\hat E_\text{MPL}$ and $E_\text{prior}$. Let us summarize the adaptive MPL-MLME strategy with the following scheme:
\begin{enumerate}
  \item Randomly choose $\rho_\text{i}^{(1)}$ as the first input state and set $l=1$.
  \begin{enumerate}
    \item Perform QPT using $\rho_\text{i}^{(l)}$ and obtain the set of frequencies $\{\nu_{l1},\ldots,\nu_{lM}\}$, $\sum_m\nu_{lm}=1$.
    \item Set $\nu=\bigcup^l_{j=1}\{\nu_{j1},\ldots,\nu_{jM}\}/\,l$.
    \item Invoke the MLME algorithm with $\nu$ and obtain $\hat E^{(l)}_\text{MLME}$.
    \item Using $E_\text{prior}$, generate a set of pairs of MPL estimators ($\hat\rho_\text{MPL}$,\,$\hat E_\text{MPL}$), where the states $\hat\rho_\text{MPL}$ were not part of the $l$ input states previously used, by iterating Eqs.~(\ref{MPLextremal}) from different, randomly chosen starting points.
    \item Set $\rho_\text{i}^{(l+1)}$ as the input state corresponding to the state estimator $\hat\rho_\text{MPL}$ such that $\mathcal{D}_\text{tr}\left(\hat E_{\text{MPL}},\hat E^{(l)}_\text{MLME}\right)$ is the largest.
  \end{enumerate}
  \item Set $l=l+1$ and repeat Steps 1(a)--1(e).\\
\end{enumerate}

\begin{figure}[htp]
  \centering
  \includegraphics[width=0.4\textwidth]{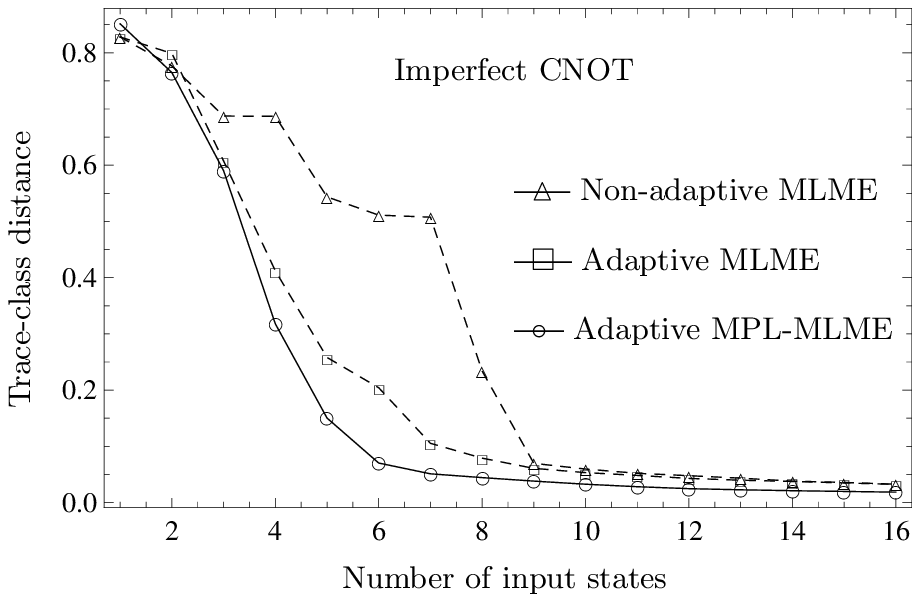}\\
  \includegraphics[width=0.4\textwidth]{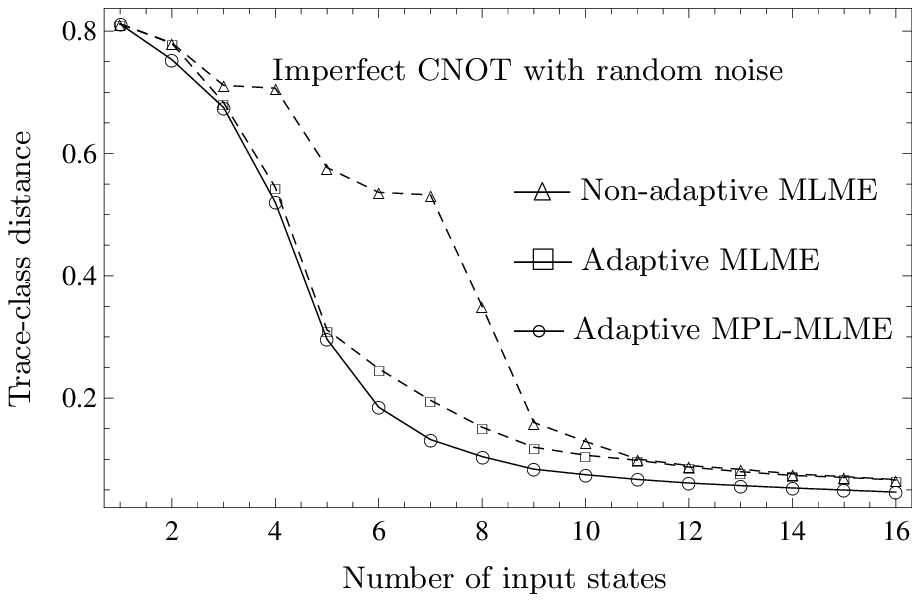}
  \caption{A comparison of three incomplete QPT schemes: the non-adaptive MLME scheme, the adaptive MLME scheme and the adaptive MPL-MLME scheme. Monte Carlo simulations are carried out on two different types of imperfect CNOT gates described in the text. Here, $N=10^4$ and an average over 50 experiments is taken to compute the trace-class distances. For both the non-adaptive as well as the adaptive MLME schemes, the 16 linearly independent input states are chosen to be tensor products of projectors of the kets $\ket{0}$, $\ket{1}$, $(\ket{0}+\ket{1})/\sqrt{2}$ and $(\ket{0}+\ket{1}\I)/\sqrt{2}$ \cite{remarks}. For all schemes, the POM outcomes are chosen to be the tensor products of qubit SIC POMs. The tomographic performance of the adaptive MPL-MLME scheme is the best out of the three. The plots show that the tomographic efficiency can be further improved by optimizing the input states over the Hilbert space instead of restricting to a fixed set of linearly independent input states, albeit the small difference in tomographic performance between the two adaptive schemes for some quantum processes.}
  \label{fig:comp}
\end{figure}

With this, let us first compare the performances of the three proposed schemes, namely the non-adaptive MLME scheme in Sec.~\ref{sec:iteralgo}, the adaptive MLME scheme in Sec.~\ref{subsec:fixed_input} and the adaptive MPL-MLME scheme. For this purpose, we consider two quantum processes, the first being an imperfect CNOT gate whose action is described by the Kraus operators
\begin{equation}
K_1=\sqrt{1-\epsilon}\,U_\text{CNOT}\quad\text{and}\quad K_2=\sqrt{\epsilon}\,.
\label{cnot_id}
\end{equation}
This first channel is a CNOT gate with probability $1-\epsilon$ and does nothing to the input states with probability $\epsilon$, an imperfect CNOT gate represented by a rank-2 Choi-Jami{\'o}{\l}kowski operator. The second process is described by the Kraus operators
\begin{equation}
K_1=\sqrt{1-\epsilon}\,U_\text{CNOT}\quad\text{and}\quad\left\{K_j=\sqrt{\epsilon}B_j\right\}^{16}_{j=2}\,,
\label{cnot_rand}
\end{equation}
where the 15 operators $B_j$ are randomly generated and satisfy the equation $\sum_jB^\dagger_jB_j=1_\mathcal{K}$. This second channel, which is represented by a full-rank matrix, is a CNOT gate with probability $1-\epsilon$ and randomly perturbs the input states with probability $\epsilon$ due to additional noise. As an example, we set $\epsilon=0.1$. Figure~\ref{fig:comp} shows the numerical results.

Next, to understand how this adaptive MPL-MLME strategy can lead to an optimization in tomographic performance, we need to know how increasing the number of input states used in AMLME QPT can affect the corresponding MLME estimators. Since we are considering only a subset of the full linearly independent input states in general, there exists a convex set of estimators $\hat E_\text{ML}$ maximizing the likelihood functional for a given set of informationally incomplete measurement data. This means that the likelihood functional possesses a plateau hovering over this convex set of estimators. As the number of input states $L$ used increases, the likelihood plateau will either remain unchanged (if no additional information about $E_\text{true}$ is gained after performing QPT with new input states) or decrease in size (if new independent information is obtained). Thus in general, the plateau will continue to shrink to a point when a full set of linearly independent input states is used.

We conjecture that the adaptive MPL-MLME strategy optimizes the rate of decrease in the size of the likelihood plateau by maximizing the normalized projected log-likelihood functional with respect to the next input state. A point of view to justify this conjecture is to interpret the maximum of the normalized log-likelihood functional $\log\left(\mathcal{L}(E)\right)/LN$ as the maximum information gain from the measurement data. When the number of copies $N$ is infinite, the data are noiseless and the resulting maximum information gain is $\sum_{lm}f_{lm}\log(f_{lm})$, which is the negative of the Shannon entropy of the measurement data. For finite $N$, the maximum information gain over the space of density operators will typically be lower than the true maximum due to the positivity constraint, especially when $E_\text{true}$ is highly rank-deficient. In this language, the MPL-MLME strategy attempts to maximize this maximum information gain as much as possible via the optimization of future input states over the entire Hilbert space of density operators, using the normalized projected log-likelihood functional as an estimate for the actual normalized log-likelihood functional describing future measurements. This is a possible explanation for the optimal decrease in the likelihood plateau size since one has maximal knowledge about the unknown $E_\text{true}$ gained with the optimized input states and so the ambiguity in the estimators is minimized.

\begin{figure}[htp]
  \centering
  \includegraphics[width=0.4\textwidth]{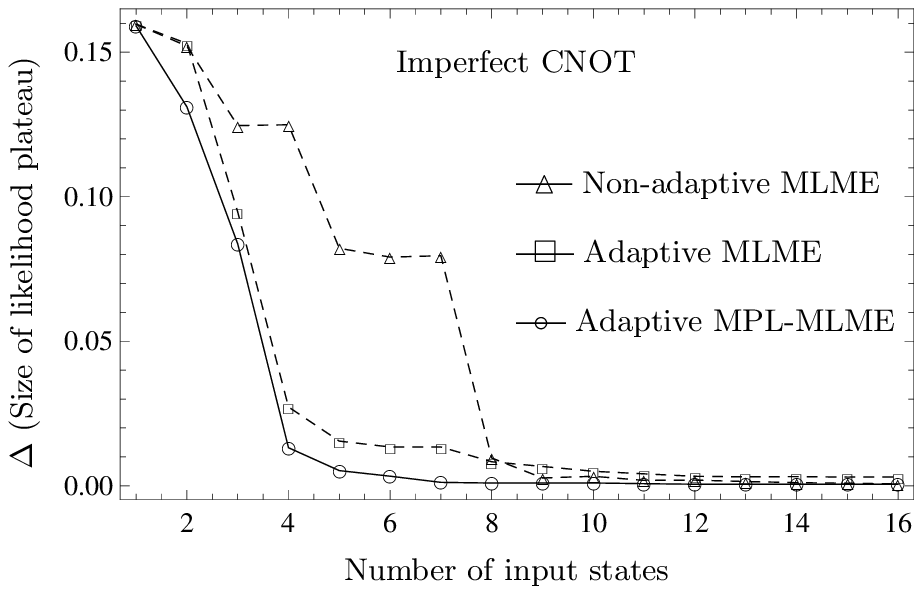}\\
  \includegraphics[width=0.4\textwidth]{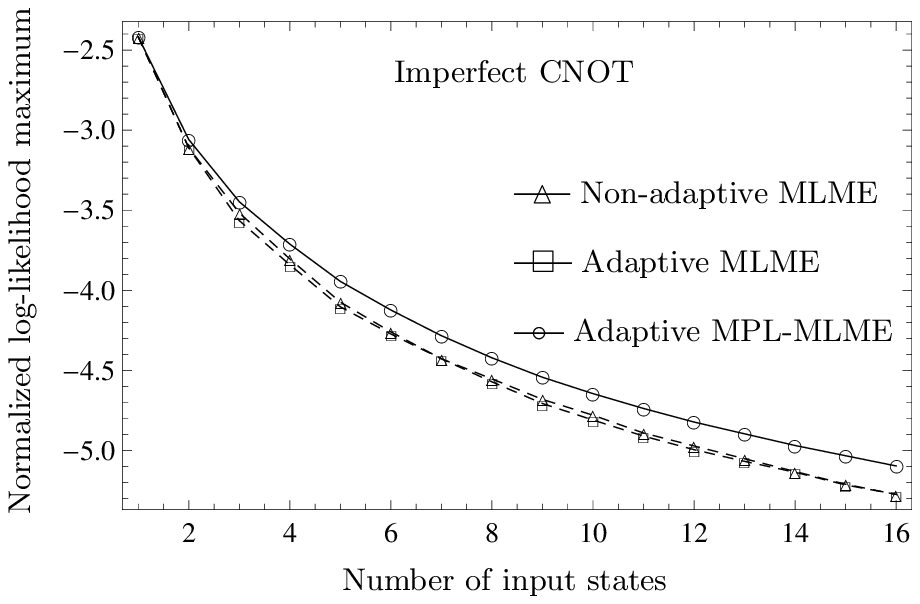}
  \caption{The dependence of the size of the likelihood plateau ($\Delta$) and the normalized log-likelihood maximum on the number of input states. The respective performances of the non-adaptive MLME scheme, the adaptive MLME scheme and the adaptive MPL-MLME scheme are computed based on noiseless measurement data for an imperfect CNOT gate with $\epsilon=0.1$. For both the non-adaptive MLME scheme and the adaptive MLME scheme, the 16 linearly independent input states are chosen to be tensor products of projectors of the kets $\ket{0}$, $\ket{1}$, $(\ket{0}+\ket{1})/\sqrt{2}$ and $(\ket{0}+\ket{1}\I)/\sqrt{2}$. For all schemes, the POM outcomes are chosen to be the tensor products of qubit SIC POMs. From the plot, the rate of decrease of $\Delta$ is the greatest with the adaptive MPL-MLME scheme. The increase in the normalized log-likelihood maxima with the adaptive MPL-MLME scheme may also be interpreted as greater maximum information gain after measurements using the optimal input states as compared to the other schemes.}
  \label{fig:plateau_height}
\end{figure}

We illustrate this point by considering the imperfect CNOT gate with $\epsilon=0.1$ described by Eq.~(\ref{cnot_id}). Since the boundary of the likelihood plateau is complicated, we shall estimate its size numerically by first generating $N_0=10^3$ ML estimators $\hat E^{(j)}_\text{ML}$ labeled with the index $j$ for a given set of measurement data. Next, in the same spirit as in numerical sampling, we can define the operator centroid
\begin{equation}
\bar E_\text{ML}=\frac{1}{N_0}\sum^{N_0}_{j=1}\hat E^{(j)}_\text{ML}
\end{equation}
for this generated set of estimators and the normalized Hilbert-Schmidt standard deviation
\begin{equation}
\Delta=\frac{1}{\ID}\sqrt{\frac{\sum^{N_0}_{j=1}{\tr{\left(\hat E^{(j)}_\text{ML}-\bar E_\text{ML}\right)^2}}}{2N_0}}\,
\end{equation}
away from the centroid. Thus, $0\leq\Delta\leq 1$. For sufficiently large $N_0$, the size of the plateau may be well approximated by the spread $\Delta$. Figure~\ref{fig:plateau_height} compares the respective performances of the the three proposed schemes by analyzing the size of the likelihood plateau and the maximum of the normalized log-likelihood functional. From Fig.~\ref{fig:plateau_height}, it is crucial to understand that $\Delta$ does not, strictly speaking, decrease monotonically with increasing height of the normalized log-likelihood functional. A counterexample is shown in the figure, that is a significant decrease in $\Delta$ for the adaptive MLME scheme as compared to the non-adaptive one with the corresponding slight decrease in its normalized log-likelihood maxima. We emphasize that what the adaptive MPL-MLME strategy exploits is the possible \emph{trend} of this behavior.

To end this part of the section, we comment that the aforementioned idea can be applied to adaptively choose the next set of POM outcomes $\Pi_j$ based on the collected measurement data. However, to perform the optimization successfully requires the solutions to more technical problems which include ensuring that the POM outcomes are linearly independent after the optimization. This project is left for future studies.

\subsection{A combination of both adaptive strategies}
\label{subsec:combination}

Let us begin this final part of the section by reviewing the non-convex feature of the MPL-MLME strategy discussed in Sec.~\ref{subsec:all_input}. The presence of multiple local-maxima estimators which are linearly independent is an important element of the MPL-MLME strategy as it provides linearly independent input states which are optimal for measurement based on the data obtained from the experiments. In general, because of the non-linearity of Eq.~(\ref{MPLextremal}), it is difficult to determine the number of such linearly independent extremal solutions for a given set of measurement data by analytical means. One can only search for as many linearly independent local-maxima estimators $\hat\rho_\text{MPL}$ as possible via numerical optimizations from different starting points within a reasonable time period.

\begin{figure}[htp]
  \centering
  \includegraphics[width=0.4\textwidth]{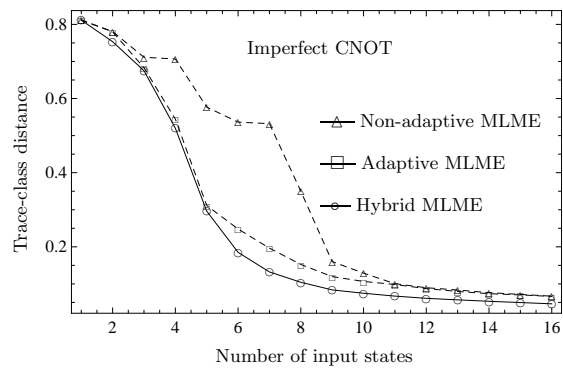}
  \caption{A comparison of three incomplete QPT schemes: the non-adaptive MLME scheme, the adaptive MLME scheme and a combination of the adaptive MPL-MLME scheme and the adaptive MLME scheme (hybrid scheme). Monte Carlo simulations are carried out on the imperfect CNOT gate with $\epsilon=0.1$. Here, $N=10^4$ and an average over 50 experiments is taken to compute the trace-class distances. For both the non-adaptive as well as the adaptive MLME schemes, the default set of 16 linearly independent input states are chosen to be tensor products of projectors of the kets $\ket{0}$, $\ket{1}$, $(\ket{0}+\ket{1})/\sqrt{2}$ and $(\ket{0}+\ket{1}\I)/\sqrt{2}$. For all schemes, a set of 16 randomly generated positive operators, which are all linearly independent of one another, are used to form the POM. For this POM, the average repetition frequency of the adaptive MPL-MLME scheme is very high after four input states are used. The first input state for all schemes is chosen to be the same separable state $\rho^{(1)}_\text{i}=\ket{00}\bra{00}$. For the third scheme, the second to the fourth input states (shaded region) are optimized using the adaptive MPL-MLME strategy and the subsequent input states are chosen via the adaptive MLME strategy using the default set of input states which excludes $\ket{00}\bra{00}$. The plot shows that the overall performance of the combined strategy is better than the adaptive MLME strategy alone.}
  \label{fig:combined}
\end{figure}

Another technical subtlety is that these local-maxima estimators tend to repeat themselves during the optimization. Hence, a local-maxima estimator which was chosen as one of the input states earlier may reappear in later optimizations. The repetition frequency strongly depends on the POM chosen to measure the output states. The examples given thus far make use of the product tetrahedron measurements as the POM and the resulting MPL optimizations give linearly independent estimators with few repetitions. This may not be the case for other types of POM. In view of this, another way of doing AMLME QPT is to use both adaptive strategies in Secs.~\ref{subsec:fixed_input} and \ref{subsec:all_input} interchangeably, the \emph{hybrid} MLME strategy. For example, one can start with the adaptive MPL-MLME strategy for tomography and when the repetition rate increases as more input states have been used, one may switch to the first adaptive MLME strategy. Figure~\ref{fig:combined} suggests that such a hybrid MLME strategy can further improve the tomographic performance as compared with the adaptive MLME strategy alone.

\section{Summary}

We have established adaptive numerical strategies to perform incomplete quantum process tomography. One may choose whichever strategy that is convenient to her to carry out tomography depending on the available types of measurement resources at hand. Each of these strategies combines the simplicity of incomplete quantum process tomography using quantum state estimation with good tomographic performances using optimization techniques. It can never be overemphasized that, although some prior information is necessary for each adaptive strategy, such information is \emph{never} used in the estimation of the unknown quantum process. Rather, the prior information is utilized to adaptively select future measurement resources, the input states in our context, based on the current measurement data, to optimize the tomographic performance. The discussions presented in this article, therefore, provide a means of obtaining estimators for the unknown quantum process using incomplete resources which are typically within reasonably good experimental precisions. These estimators are statistically meaningful in that they are least-biased with respect to a set of informationally incomplete measurement data and are hence suitable for partial characterization of quantum processes. This is in contrast with the standard quantum process tomography which generally requires a huge amount of informationally complete measurement resources.

\begin{acknowledgements}
We would like to thank Guo Chuan Thiang and Huangjun Zhu for fruitful and stimulating discussions. This work is supported by the NUS Graduate School for Integrative Sciences and Engineering and the Centre for Quantum Technologies which is a Research Centre of Excellence funded by Ministry of Education and National Research Foundation of Singapore, as well as the Czech Ministry of Education, Project MSM6198959213, and the Czech Ministry of Industry and Trade, Project FR-TI1/364.
\end{acknowledgements}


\begin{thebibliography}{99}

\bibitem{qpt0}
Throughout this article, the words ``quantum process'' and ``quantum channel'' will be used interchangeably.

\bibitem{qpt1}
M. Mohseni, A. T. Rezakhani, and D. A. Lidar, \pra\,\textbf{77}, 032322 (2008).

\bibitem{qpt2}
J. L. O'Brien, G. J. Pryde, A. Gilchrist, D. F.V. James, N. K. Langford, T. C. Ralph, and A. G. White, \prl\,\textbf{93}, 8 (2004).

\bibitem{qpt3}
J. F. Poyatos, J. I. Cirac, and P. Zoller, \prl\,\textbf{78}, 2 (1997).

\bibitem{nielsenchuang}
M. A. Nielsen and I. L. Chuang, \emph{Quantum Computation and
Quantum Information}, Cambridge University Press, Cambridge, (2000).

\bibitem{choijam}
M. Choi, Linear Algebr. Appl. \textbf{10}, 285 (1975); A. Jami{\'o}{\l}kowski, Rep. Math. Phys. \textbf{3}, 275 (1972).

\bibitem{rehacek1}
M. Paris, and J. {\v R}eh{\'a}{\v c}ek, \emph{Lecture Notes in Physics \textbf{649} - Quantum State Estimation}, Springer, Berlin Heidelberg (2004); J. Fiur{\'a}{\v s}ek and Z. Hradil \pra\,\textbf{63} 020101(R) (2001).

\bibitem{csqpt}
S. Rahimi-Keshari, A. Scherer, A. Mann, A. T. Rezakhani, A. I. Lvovsky, and B. C. Sanders, New J. Phys.\,\textbf{13}, 013006 (2011); M. Lobino, D. Korystov, C. Kupchak, E. Figueroa, B. C. Sanders, and A. I. Lvovsky, Science\,\textbf{322}, 563 (2008).

\bibitem{jaynes}
E. T. Jaynes, Phys. Rev.\,\textbf{106}, 620 (1957), Phys. Rev.\,\textbf{108}, 171 (1957).

\bibitem{iqpt}
M. Ziman, \pra\,\textbf{78}, 032118 (2008).

\bibitem{mlme}
Y. S. Teo, H. Zhu, B.-G. Englert, J. {\v R}eh{\'a}{\v c}ek, and Z. Hradil, \prl\,\textbf{107}, 020404 (2011).

\bibitem{sicpom}
A. J. Scott, and M. Grassl, J. Math. Phys.\,\textbf{51}, 042203 (2010); A. Ling, A. Lamas-Linares, and C. Kurtsiefer, \texttt{eprint} 0807.0991 (2008); J. {\v R}eh{\'a}{\v c}ek, B.-G. Englert, and D. Kaszlikowski \pra\,\textbf{70}, 052321 (2004).

\bibitem{iontrap}
M. Riebe, K. Kim, P. Schindler, T. Monz, P. O. Schmidt, T. K. K{\"o}rber, W. H{\"a}nsel,
H. H{\"a}ffner, C. F. Roos, and R. Blatt, \prl\,\textbf{97}, 220407 (2006).

\bibitem{mce1}
D. Jurafsky, and J. H. Martin, \emph{Speech and Language Processing}, Prentice Hall (2009).

\bibitem{mce2}
P. A. Est\'evez, C. J. Figueroa, and K. Saito, Proc. IEEE IJCNN\,\textbf{5}, 2724 (2005).

\bibitem{remarks}
This set of input states, taken from Ref.~\cite{iontrap}, is just one of the many possible choices one can use in quantum process tomography. It is important to understand that this set is by \emph{no means} sanctioned to be the ``standard'' set of input states. Rather, these are four states of the six projectors of the standard six-outcome POM, but \textbf{any} four of the six states will serve the purpose equally well.

\end{thebibliography}
\end{document}